\documentclass{moriond}
\usepackage{amsmath}
\usepackage{amssymb}
\usepackage{xspace}
\usepackage{mcite}

\def\nubar{\overline\nu}
\def\numu{\nu_\mu}
\def\numubar{\overline\nu_\mu}
\def\nue{\nu_e}
\def\nuebar{\overline\nu_e}
\def\unit#1{\,\mathrm{#1}}
\def\abs#1{\left|#1\right|}
\def\ue{\mathrm{e}}
\def\ui{\mathrm{i}}
\def\POD{P$\emptyset{}$D\xspace}

\def\be{\begin{equation}}
\def\ee{\end{equation}}
\def\bea{\begin{eqnarray}}
\def\eea{\end{eqnarray}}

\newcommand{\Photo}{}

\begin{document}
\vspace*{4cm}
\title{Latest activities and results from T2K}

\author{R.~P.~Litchfield, on behalf of the T2K collaboration}

\address{University of Glasgow, School of Physics and Astronomy,\\ Glasgow, G12 8QQ, United Kingdom}

\maketitle\abstracts{
The T2K neutrino oscillation experiment has gone through a period of renewal over the last couple of years, with several upgrades designed to improve sensitivity to leptonic CP violation. The change that most affects future analyses is the ND280 upgrade, which will constrain interaction models in new ways.  At the same time the analysis continues to be developed, both by combining with the Super-Kamiokande atmospheric measurements and with refinements to the T2K-only analysis.  This proceedings describes the highlights of these analyses, and the status of the various T2K upgrades.}

\section{Introduction and physics goals of T2K}

The Tokai to Kamioka (T2K) experiment is a long-baseline neutrino experiment studying the $\numu\rightarrow\numu$ and $\nue\rightarrow\nue$ oscillation channels (and their antimatter counterparts) with a baseline $L = 295\unit{km}$, and a neutrino beam that peaks around $E = 0.6\unit{GeV}$.  T2K has been taking data since 2009, using the preexisting Super-Kamiokande (SK) detector and a suite of near detectors installed at $280\unit{m}$ from the pion production target, which receives fast extracted spills of $30\unit{GeV}$ protons from the Main Ring of the Japan Proton Accelerator Research Center (J-PARC).

The original physics goal of T2K was to observe the $\numu\rightarrow\nue$ oscillation for the first time, which would indicate that the $U_{e3}$ element of the PMNS mixing matrix (commonly parameterised as $\sin\theta_{13}\, \ue^{-\ui \delta_{CP}}$) was non-zero, or to place limits on its size a factor of 20 below what was known in 2009.  It turned out\cite{T2K:firstnue, DC2012,*DB2012,*RENO2012} that $\abs{U_{e3}}$ was very close to the existing limit, which allowed T2K to discover the $\numu\rightarrow\nue$ oscillation very early in its operation\cite{T2K:nuediscovery}.  The physics activity of T2K subsequently switched to making precision measurements of the four oscillation channels, with particular focus on the possibility of CP violation in the lepton sector. This is closely associated with the complex phase parameter $\delta_{CP}$, and is observed in the experiment as a difference between the oscillation probabilities $P(\numu\rightarrow\nue)$ and $P(\numubar\rightarrow\nuebar)$. 

\begin{figure}
    \centering
    \hfill
    \parbox{0.25\textwidth}{\centering 
        \includegraphics[width=0.21\columnwidth]{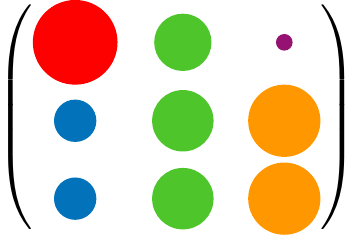}
        \scriptsize{Normal Ordering\\ ~}\\
        \includegraphics[width=0.21\columnwidth]{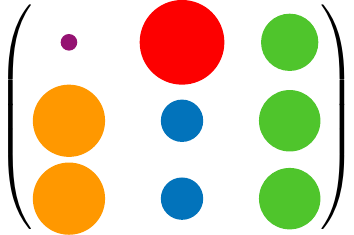} 
        \scriptsize{Inverted Ordering\\ ~}\\
    }
    \hfill
    \parbox{0.25\textwidth}{\centering 
        \includegraphics[width=0.21\columnwidth]{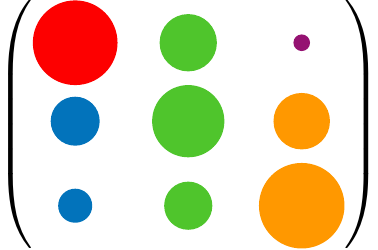} \\ \scriptsize{Lower Octant \\ ~}\\
        \includegraphics[width=0.21\columnwidth]{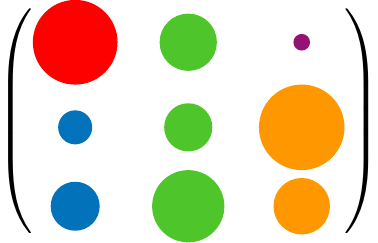} \\
        \scriptsize{Upper Octant \\ ~}\\
    }
    \hfill
    \parbox{0.25\textwidth}{\centering 
    \includegraphics[width=0.25\columnwidth]{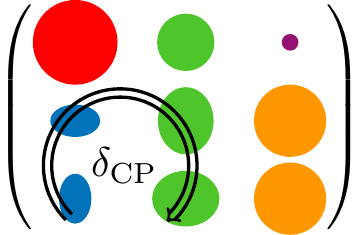} \\ \scriptsize{Variation with $\delta_\mathrm{CP}$ \\ ~}
    }
    \hfill \vphantom{.}\\
    \caption{Unresolved questions about neutrinos and leptonic mixing:  There are discrete options for the mass ordering and octant, and in only one of them (NO, $\theta_{23} < \frac{\pi}{4}$) would the leptons retain a generational (i.e. leading diagonal) structure.  The matrix may also have a unitary form, leading to CP violation.}
    \label{fig:PMNS_questions}
\end{figure}

Regardless of the parametrisation, the most important questions about the mixing of leptons can be understood in terms of the magnitude-squared of the elements of the $3\times3$ PMNS mixing matrix, as shown in Figure~\ref{fig:PMNS_questions}.  The rows of the matrix correspond to the flavour states $\nu_\alpha = \{\nue, \numu, \nu_\tau\}$, while the columns are labelled $\nu_i = \{\nu_1, \nu_2, \nu_3\}$, ordered by decreasing overlap with the electron flavour (i.e. the size of the top row elements).  

The first important question is whether this ordering also corresponds to increasing neutrino mass, which is termed \emph{normal ordering} (NO) by analogy with the quarks. The alternative is \emph{inverted ordering} (IO), where $\nu_3$ is actually the lightest neutrino.  The second important question is also discrete: we do not know which of $\abs{U_{\mu3}^2}$ and $\abs{U_{\tau3}^2}$ is greater. This is termed the \emph{octant degeneracy}, as it corresponds to whether or not the angle $\theta_{23} <\pi/4$ in the standard parametrisation. 

The final important question relates to whether the PMNS matrix can be expressed as a purely real transformation (a 3D rotation), or whether there is an irreducible complex phase, $\delta_{CP}$, making the transformation unitary. This extra degree of freedom is manifest as an extra mixing of the lower left four elements.  If the matrix is not expressible as a real rotation the relative phases introduced take opposite sign for neutrinos and antineutrinos, resulting in a difference between $P(\numu\rightarrow\nue)$ and $P(\numubar\rightarrow\nuebar)$.  In T2K the main effect of this difference is on the overall $\nue$-like event rates for neutrinos and antineutrinos.  However, interpretation of such a signal can be ambiguous for many combinations of oscillation parameters, as the mass ordering has a similar effect on the total rates: In T2K NO enhances $P(\numu\rightarrow\nue)$ and reduces $P(\numubar\rightarrow\nuebar)$, and vice versa for IO.     

\subsection{Recent T2K activity}
The T2K experiment itself took data in essentially its original configuration until April 2021, and achieving a maximum power of $523\unit{kW}$. In 2021 and 2022, the J-PARC main ring was upgraded to allow a faster repetition rate, from $2.48\unit{s}$ to $1.36\unit{s}$.  In 2023 the experiment restarted and on $25^\mathrm{th}$ December a new record of $760\unit{kW}$ was set, shortly after receiving governmental permission to exceed the original $750\unit{kW}$ maximum beam power.\!\footnote{In June 2024 a new record of $800\unit{kW}$ was achieved.}  The power supplies that energise the focusing horns were also upgraded, allowing the horn currents to be raised from $250\unit{kA}$ to $320\unit{kA}$.  This increases the strength of magnetic focusing, which improves the number of correct-sign $\numu$ in the flux peak by about 10\%.

During 2022, the SK detector was also loaded 
to a $0.03\%$ concentration of Gadolinium. This is to detect neutrons, and can in principle improve identification of $\overline\nu_\ell + p \rightarrow \ell^+ + n$  interactions.  However this is mostly intended for low-energy neutrinos, and is not so important for neutrino-mode running in T2K.  The off-axis near detector is also seeing substantial change, in the form of \emph{ND280 upgrade}. This entirely replaces the upstream section of the detector and is currently being commissioned.

The remainder of this proceedings will describe recent activity by the collaboration: the latest T2K analysis published using the pre-upgrade data; an analysis of beam and atmospheric neutrinos performed jointly with the Super-Kamiokande collaboration; and some details about the ND280 upgrade.\footnote{T2K has also been working on a joint analysis with the NO$\nu$A experiment, based on both experiment's existing 2020 analyses. This was reported in the following talk by M.\,Sanchez.}

\section{T2K-only analysis}
The T2K analysis presented here was published in 2023\cite{T2K_OA2023}, and includes SK data up to 2020 and a new $\numu \textrm{CC}1\pi$ sample at SK. This represents a modest 15\% increase in the statistical power of the oscillation measurement itself. There is a much more substantial 106\% increase in the amount of near detector data used compared to the previous analysis\cite{T2K_OA2021}.  This large increase was made possible by a number of long term efforts to correct readout issues in recent runs that would have reduce the quality of key selection cuts and degraded the purity of many samples.  The significantly increased data set also makes it worthwhile to subdivide the data taken in $\nubar$-mode by the number of pions observed---as is done for the $\nu$-mode---and to bin this data more finely in energy than previous analyses.

Modelling of the flux has been significantly improved (from 9\% to 5\% uncertainty in the peak) using data from NA61/SHINE\,\cite{NA61_replica,*NA61_thin}, and a significantly expanded cross section model 
.  Although this final change is an improvement, the extra degrees of freedom it provides allow for larger model variation, even with the stronger ND280 constraint. This leads to correspondingly larger intervals on oscillation parameters, particularly the mass splitting parameter $\Delta m^2_{32}$. 

Finally, constraints on oscillation parameters are affected by updating the external `reactor constraint' on $\theta_{13}$, to incorporate the most recent measurements by the RENO and Daya Bay experiments.  In the previous analysis the constraint  $\sin^2(\theta_{13}) = 2.12\pm0.08$, while in this update a world average value of $\sin^2(\theta_{13}) = 2.18\pm0.07$ is used\cite{PDG_2019}. For the region of parameter space T2K data prefers, a larger value of $\theta_{13}$ enhances the number of electron-type (anti)neutrinos seen and tends to enlarge the allowed regions. 
\begin{figure}[t]
    \parbox{0.55\textwidth}{
    \centering
    \includegraphics[width=0.55\textwidth, trim={0 0 0 180}, clip]{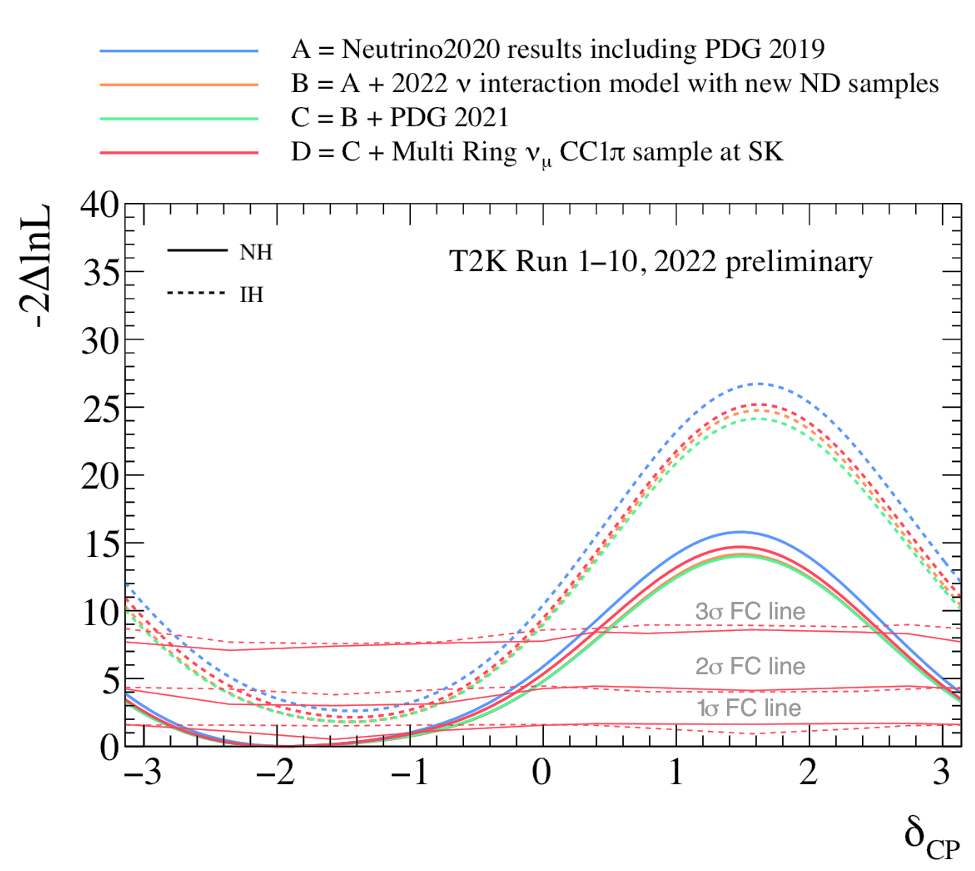}
    }
    \parbox{0.44\textwidth}{
    \centering
    \vspace{-6ex}
    \includegraphics[width=0.44\textwidth, trim={85 685 35 0}, clip]{figs/T2K_only_dcp.png}
    \\[9ex]
    \begin{tabular}{rc|cc}
         \multicolumn{2}{l|}{T2K posterior}& $\theta_{23} < \frac{\pi}{4}$ & $\theta_{23} > \frac{\pi}{4}$\\   \multicolumn{2}{l|}{probability} & 0.23 & 0.77\\[0.3ex]
         \hline & \\[-2.4ex]
         $\Delta m^2_{32} > 0$& 0.80 & 0.19 &0.61 \\ 
         $\Delta m^2_{32} < 0$& 0.20 & 0.04 &0.16 
    \end{tabular}
    }\\
    \caption{[L] Evolution of confidence intervals on the CP phase. The new interaction model and change to the reactor constraint both widen the intervals, slightly more than the additional data helps constrain them.\quad[R]~Posterior probabilities for the mass ordering ($\mathrm{sign}[\Delta m^2_{32}]$), and octant ($\mathrm{sign}[\theta_{32} - \frac{\pi}{4}]$) degeneracies, obtained using T2K data only.}
    \label{fig:T2Konly_dcp}
\end{figure}

As in previous analyses, the parameterised flux and interaction models are first constrained with external data, then fit to the Near Detector data to produce a ND posterior probability distribution. The ensemble of models this represents is then used as a the basis for a marginal likelihood function of the SK data, which also includes the oscillation parameters.  The marginal likelihood is used in both frequentist and Baysian analysis to derive the oscillation parameters.  Figure~\ref{fig:T2Konly_dcp}[L] shows frequentist confidence intervals on the parameter $\delta_{CP}$, and the Baysian credible intervals are very similar.  All CP-conserving solutions are outside the 90\% interval, but the $(\mathrm{NO},\delta_{CP}=\pi)$ point is within the $2\sigma$ confidence interval.  

The other two important questions (the mass ordering and octant degeneracy) are commonly expressed as binary choices so are conveniently addressed in a Bayesian framework, where we initially consider all four combinations to have equal prior probability.  T2K data favours both the upper octant and normal mass ordering solutions, as shown in Figure~\ref{fig:T2Konly_dcp}[R] but the preference is rather weak, only around a 4:1 odds ratio in either case.

\section{T2K+SK analysis}
Analysis of atmospheric neutrino data predates the construction of T2K and is carried out independently by the Super-Kamiokande collaboration.  However SK alone lacks the high resolution ND280 data that can constrain cross-section models through high-statistics measurement of exclusive channels. SK atmospheric analyses also cannot know the incident neutrino direction (and therefore the exact energy and baseline) which also means that T2K is better at determining a precise value of $\delta_{CP}$ and the mass splitting $\abs{\Delta m^2_{32}}$. 

On the other hand, SK atmospheric data gives access to the MSW resonance, which is a clear indicator of the mass ordering.  In Earth's crust this occurs above $10\unit{GeV}$, well away from the beam energy of existing beam-based experiments, but in Earth's lower mantle and core the higher density lowers the resonance to around $7$ and $3\unit{GeV}$ respectively, within the flux of multi-GeV neutrinos that SK can observe. The resonance will occur for the more dominant neutrino channel if the mass ordering is NO but not for IO.  The wider energy spectrum also helps distinguish between the two CP conserving values of $\delta_{CP} = \{0,\pi\}$, which are almost degenerate in T2K-only analyses.
\begin{figure}
    \centering
    \includegraphics[height=0.3\textwidth]{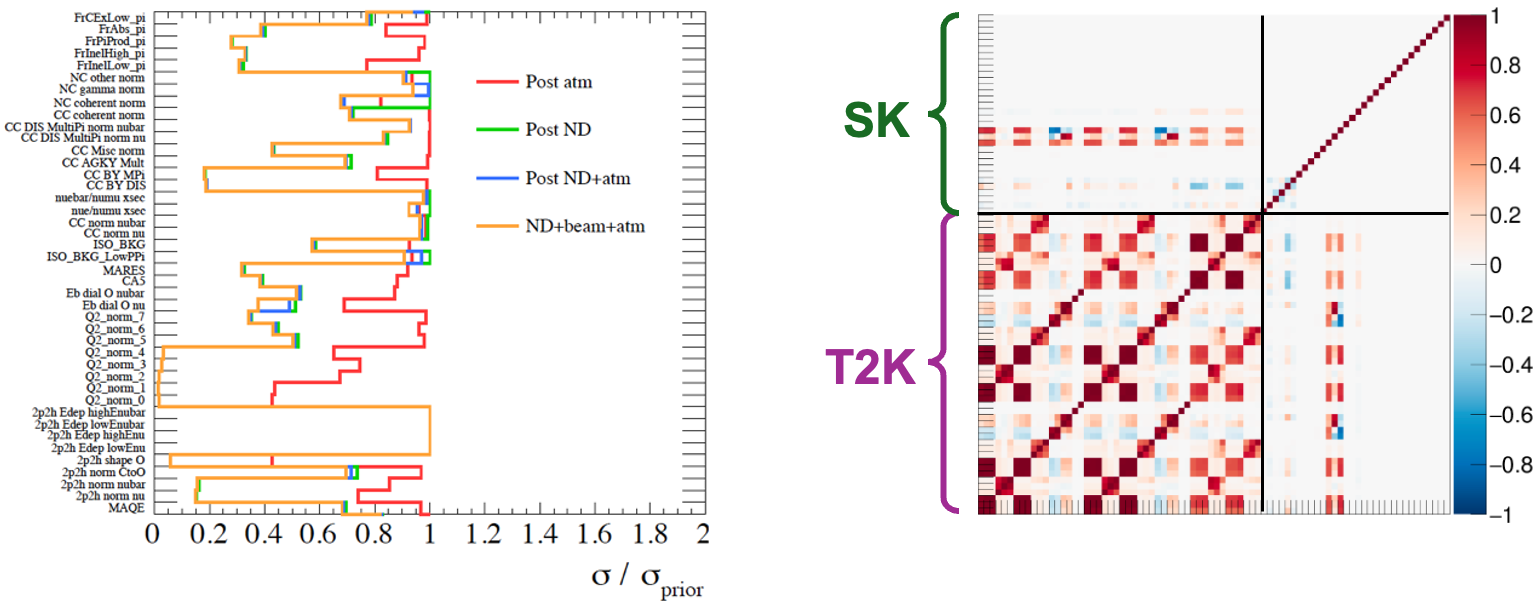}
    \caption{[L] Cross section systematic uncertainties, showing how T2K's ND280 (green) can constrain the model more strongly than SK alone (red).\quad [R] Correlation matrix showing how SK detector parameters are correlated between SK and T2K samples.}
    \label{fig:T2KSK_correaltion}
\end{figure}

The combined T2K+SK analysis uses the complimentary strengths of the two experiments, and is a much deeper integration than the T2K+NOvA joint analysis, as shown in Figure~\ref{fig:T2KSK_correaltion}.  The two experiments share the SK detector, so the associated  uncertainties are strongly correlated (although not entirely, because of the different flux).  The interaction model is also shared, which is what allows T2K's ND280 data to constrain SK's cross-section uncertainties.

The joint analysis is based on T2K's older 2020 analysis\cite{T2K_OA2021}, and the results were recently made public at a KEK/JPARC seminar\cite{T2KSK_seminar}.  Since the talk a preprint\cite{T2KSK_preprint} of the analysis has become available.

\begin{figure}
    \centering
    \parbox{0.53\textwidth}{
    \centering
    \includegraphics[width=0.53\textwidth]{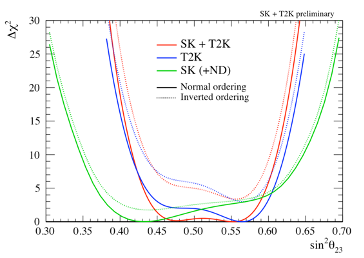}\\[4ex]
    \begin{tabular}{rc|cc}
         \multicolumn{2}{l|}{T2K+SK posterior}& $\theta_{23} < \frac{\pi}{4}$ & $\theta_{23} > \frac{\pi}{4}$\\
         \multicolumn{2}{l|}{probability} & 0.39 & 0.61\\[0.3ex]
         \hline & \\[-2.4ex]
         $\Delta m^2_{32} > 0$& 0.90 & 0.37 &0.53 \\ 
         $\Delta m^2_{32} < 0$& 0.10 & 0.02 &0.08
    \end{tabular}
    }
    \hfill
    \parbox{0.45\textwidth}{
    \centering
    \includegraphics[width=0.45\textwidth]{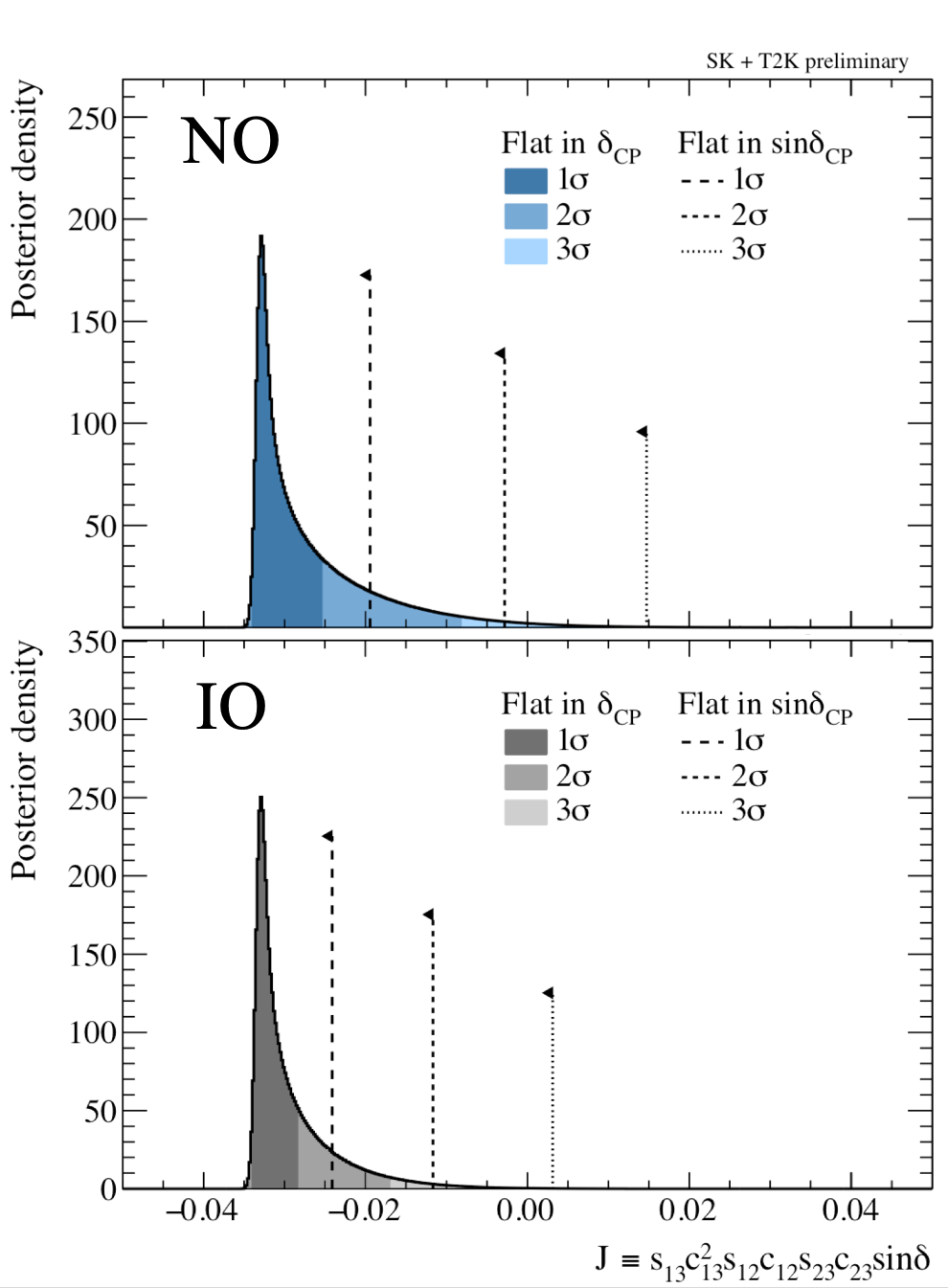}
    }
    \caption{[L] T2K and SK data have different preferences for the octant, but both prefer NO. This is reflected in how the posterior probabilities change compared to the T2K-only analysis.\quad[R] Credible intervals for $J_{CP}$ from analysis by T2K+SK. In this case the CP conserving value of 0 is just outside the $2\sigma$ interval.}
    \label{fig:T2KSK_results}
\end{figure}

An interesting aspect of the joint analysis is that the two experiments have different preferences for the octant degeneracy, and similar sensitivity, as shown in Figure~\ref{fig:T2KSK_results}[L].  However both experiments prefer the NO solution, and the posterior probability is stronger than T2K alone, with an odds ratio of 9:1 in favour of NO.  The conclusion for the CP phase is not much changed in substance, and is expressed here as a Baysian credible interval\footnote{At the time of the talk frequentist intervals were in preparation, but not finalised.} on the parametrisation-invariant measure
\[
    J_{CP} = \mathfrak{Im}\left[U_{e2}U^*_{\mu2}U^*_{e1}U_{\mu1}\right] = \tfrac{1}{8}\sin(2\theta_{23})\sin(2\theta_{12})\sin(2\theta_{13})\cos\theta_{13}\sin\delta.
\]
This is becoming increasingly popular as it is linearly related to to the empirical CP signal---the difference between the neutrino and antineutrino oscillation probabilities---but the  collaborations also continue to produce intervals calculated on $\delta_{CP}$.  The CP-conserving $J_{CP} = 0$ point is disfavoured at about the $2\sigma$ level, but whether it is outside or inside that threshold depends on the choice of presentation, and also on whether the prior is uniform in $\delta_{CP}$ or in $\sin(\delta_{CP})$.  The actual difference between these measures is small though, and the different interpretations should be seen as an artefact of the `inside/outside' framing.  Because T2K dominates the joint CP measurement, the CP conserving point is less favoured if IO is assumed, as is seen in T2K-only analyses.

\section{The ND280 Upgrade}
The original ND280 included the \POD (Pi-zero Detector), designed for the possibility that the $\nue$ appearance was an order of magnitude lower than observed and T2K's sensitivity would be limited by the the background from neutral current $\pi^0$.  Since $\abs{U_{e3}}$ is so large, this background is not significant, and measurements are more limited by the angular acceptance of the ND280.  

The ND280 upgrade replaces the majority of the \POD with a suite of detectors intended to broaden the acceptance to include more high-angle tracks, and to have better resolution of this interaction vertex.  Installation work began in 2022 with removal of the \POD, and the upgrade detectors took first beam data in August of 2023, with sections of all subsystems installed.\!\footnote{Since the talk all subsystems have been fully installed and taken data with the JPARC beam.}  About one sixth of the the original \POD remains as part of the cosmic ray trigger, and to act as active shielding of tracks entering the front of the detector.

\begin{figure}
    \centering
    \parbox{0.49\textwidth}{
    \centering
    \includegraphics[width=0.49\textwidth]{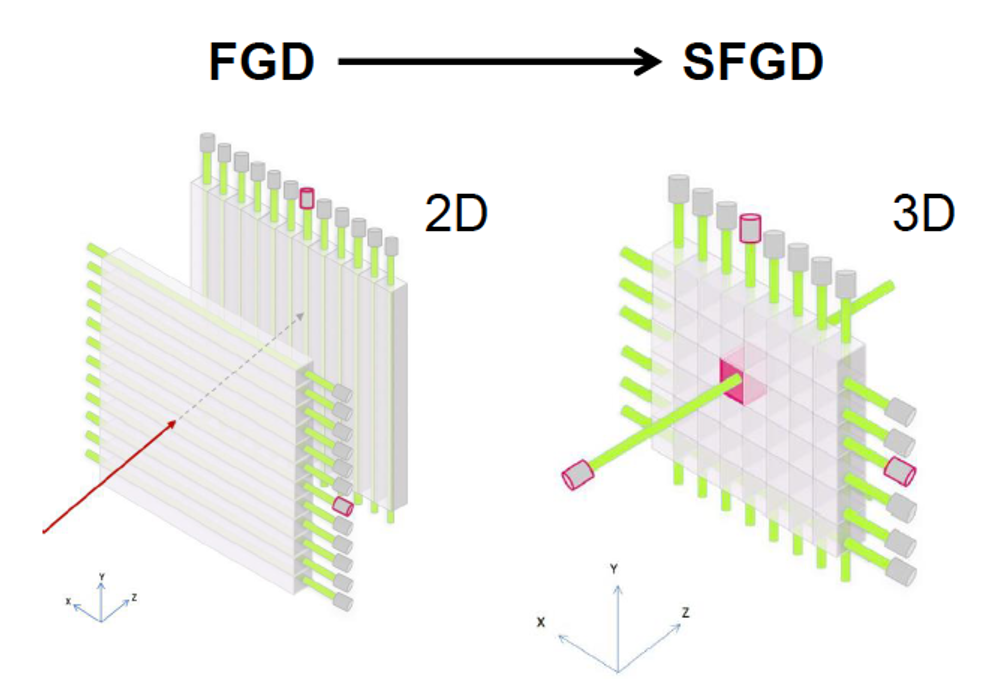}
    }
    \hfill
    \parbox{0.49\textwidth}{
    \centering
    \includegraphics[width=0.49\textwidth]{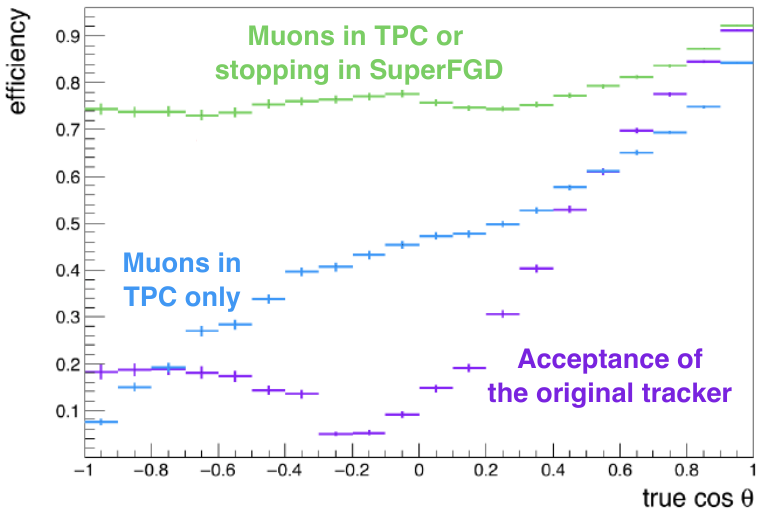}
    }
    \caption{[L] By having 3D fibre readout in every $(1\unit{cm})^3$ cube rather than alternating  $1\unit{cm}$ $x$--$y$ projections the tracking threshold of the SFGD is half that of the original FGD.\quad[R] Reconstruction acceptance of (muon) tracks in the upgrade and original tracker, showing how much acceptance improves $\gtrsim60^\circ$ from the forward direction. }
    \label{fig:upgrade_tech}
\end{figure}

The upgrade detectors use the same basic configuration as the original ND280 tracker: a fine-grained target detector made of scintillator, sandwiched between TPCs for tracking and particle ID.  But in the upgrade, the sandwich is arranged horizontally rather than vertically, so that tracks emerging at high angles (in the vertical plane) quickly enter the TPCs and are much better reconstructed.  A high resolution time-of-flight detector is also installed around the main upgrade detectors to help determine the track direction unambiguously. 
The Super-FGD target detector has a novel cube structure with fibres spaced every $1\unit{cm}$ to make 3 2D projections, as shown in Figure~\ref{fig:upgrade_tech}[L].
The fact the detector does not need alternating $x$--$y$ planes means the minimum 3D line segment is just $\sim2\unit{cm}$, rather than the $\sim4\unit{cm}$ required in the original FGD1.  And becasue the detector is also read out with longitudinally-oriented fibres it is much more effective at reconstructing high angle tracks, as there will always be at least two readout planes with useful projections.

The effect of these improvements is shown in Figure~\ref{fig:upgrade_tech}[R], and is particularly important for particles ejected perpendicularly to the beam direction.  High-angle muons correspond to interactions with large $Q^2$ and are sensitive to different regions of the nucleon (and nuclear) structure.  This extends ND280's ability to measure the neutrino--nucleus interactions, and reduces modelling dependencies when extrapolating to SK, which is has a much more isotropic acceptance.  

The full ND280 upgrade will be in operation by this summer, and with the upgraded beam a significant novel data set should become available relatively quickly. 

\end{document}